\documentstyle[aps,twocolumn]{revtex}

\begin{document}
\author{J. Bobroff$^{1}$, W.A. MacFarlane$^{1}$, H. Alloul$^{1}$, P.\ Mendels$^{1}$,
N.\ Blanchard$^{1}$, G.\ Collin$^{2}$, J.-F.\ Marucco$^{3}$}
\title{Spinless impurities in high T$_{c}$ cuprates: Kondo-like behavior}
\address{$^{1}$Laboratoire de Physique des Solides, UMR 8502, Universit\'{e}\\
Paris-Sud, 91405 Orsay, France\\
$^{2}$LLB, CE-Saclay, CEA-CNRS, 91191 Gif sur Yvette, France\\
$^{3}$LEMHE, UMR 8647, Universit\'{e} Paris-Sud, 91405 Orsay, France}
\date{\today}


\twocolumn[\hsize\textwidth\columnwidth\hsize\csname@twocolumnfalse\endcsname
\maketitle

\begin{abstract}
We compare the effects of in-plane non magnetic Li$^{+}$ 
and Zn$^{2+}$ impurities on the normal state of high-T$_{c}$ cuprates.
\ $^{89}$Y\ NMR shows that the extra hole introduced by Li 
is not localized in its vicinity. 
 The T$_{c}$ depression and induced moments on near neighbour Cu sites
of Zn or Li are found identical. These effects of spinless impurities
establish the major influence of the spin perturbation with respect to the charge 
defect. The susceptibility of the induced moment measured by $^{7}$Li NMR
 displays a 1/(T+$\Theta $) behavior. $\Theta $ increases with doping up to
about 200 K in the overdoped regime.  We attribute this to a "Kondo like" effect.
\end{abstract}

]

Increasingly, impurities are used to probe the magnetic properties of
correlated systems. For instance, in cuprates, substitution of the Cu sites
of the CuO$_{2}$ planes directly reveals the existence of magnetic
correlations in the planes and probes their interplay with
superconductivity. In particular, Zn$^{2+}$ substitution has been studied
thoroughly because it was unexpectedly found to strongly affect both the
normal and superconducting states. Above T$_{c}$, in a metallic picture, Zn$%
^{2+}$ should only weakly affect both magnetism and transport properties;
the former because it is a spinless impurity and the latter because it has
the same charge as Cu$^{2+}$. In contrast, Zn acts as a very strong
scattering center \cite{ChienPRL}. The fact that T$_{c}$ is depressed by
this scattering is primarily a consequence of the now well established
d-wave anisotropy of the superconducting order parameter \cite{Balian}.
Furthermore, Zn induces local magnetic moments on its near neighbor (n.n.)
coppers, as shown by NMR \cite{MahajanPRL94} and macroscopic SQUID
measurements \cite{Mendels99}. Zn, as a spin vacancy, creates indeed a
perturbation of the local antiferromagnetic correlations, as also observed
in undoped low dimensional spin chains or ladders \cite{AzumaPRB97}. Such
effects were anticipated on theoretical grounds \cite{finkelstein}\cite
{PoilblancPRL94}. However, until now no experiment could clearly expose the
relation between the magnetic correlations and the scattering effects on T$%
_{c}$. Another interesting problem is the evolution of these anomalies with
hole doping.\ Recent macroscopic experiments showed that the local moment
susceptibility falls rapidly, though it still exists at optimal doping \cite
{Mendels99}. Such local moments have also been found in Al$^{3+}$
substituted LaSrCuO at optimal doping, despite some qualitative differences
with Zn \cite{IshidaPRL96}.\ In this later work, NMR of $^{27}$Al itself was
used to probe locally the susceptibility of its n.n. Cu sites. However, no
experiment has yet been dedicated to probing the evolution of this moment
into the overdoped regime. Such an experiment should help to clarify wether
the cuprates exhibit an uncorrelated Fermi Liquid behavior at high doping.

In order to address both problems, we have undertaken a study of Li in
YBaCuO which substitutes within the CuO$_{2}$ planes \cite{Sauv}. Li$^{+}$
is not magnetic like Zn$^{2+}$ but has a different valence. Comparing the
local magnetism and the effect of T$_{c}$ between Li$^{+}$ and Zn$^{2+}$
will elucidate the respective roles of charge and spin in the impurity
response of the cuprates.\ In particular, is the additional hole of Li$^{+}$
trapped in the vicinity of Li$^{+}$, as found in the undoped La$_{2}$CuO$%
_{4} $ \cite{KastnerPRB88} ? We present here $^{89}$Y\ NMR measurements
which show that the local magnetic environment of Li is identical to that of
Zn and establish both the presence of induced moments and the absence of
localized hole. The effects on T$_{c}$ are subsequently shown to be the same
for Li and Zn. In the second part, we report $^{7}$Li NMR measurements which
enable us to make a detailed study of the doping dependence of these
moments. Indeed $^{7}$Li, in contrast with $^{57}$Zn, is a sensitive NMR
probe which has already been detected in YBa$_{2}$Cu$_{3}$O$_{7}$ \cite{Sauv}%
.\ Like $^{27}$Al NMR in La$_{2}$SrCuO$_{4}$, $^{7}$Li\ NMR provides the
opportunity to measure the susceptibility of the environment of Li with
unprecedented accuracy. We present $^{7}$Li NMR data for a wide range of
concentrations and hole dopings, including in particular, for the first
time, the overdoped regime via simultaneous Ca substitution on the Y site.\
The local moment susceptibility is found to evolve from a Curie to a 1/(T+$%
\Theta $) law with increasing doping.\ Moreover, $\Theta $ is found to be
independent of impurity concentration - behavior strongly reminiscent of the
Kondo effect.

Li substituted samples YBa$_{2}$(Cu$_{1-x_{n}}$Li$_{x_{n}}$)$_{3}$O$_{6+y}$%
were prepared by solid state reaction of Y$_{2}$O$_{3}$, BaO$_{2}$, CuO, and
Li$_{2}$CO$_{3}$ with nominal Li concentration $x_{n}$. The samples were
annealed and ground many times to eliminate spurious phases. X-Ray
diffraction was performed to ensure their quality and purity. Deoxidation
was performed at fixed temperatures ranging from 340$%
{{}^\circ}%
$C to 450$%
{{}^\circ}%
$C under vacuum. As explained later, we used $^{89}$Y NMR to obtain a
reliable evaluation of the doping level. The superconducting transition was
measured in a SQUID magnetometer in field cooled experiments. T$_{c}$was
determined by extrapolation to $\chi =0$ of the Meissner transition (not as
the onset of diamagnetism). The powdered crystallites were c-axis aligned in
an epoxy matrix in a $\sim 7$ Tesla magnetic field. NMR measurements were
performed in a homemade spectrometer with a field H$_{0}$=7.5\ Tesla using
Fourier Transform spectroscopy. Spin echos were obtained using a $\pi
/2-\tau -\pi $ sequence. The frequency shifts for $^{89}$Y and $^{7}$Li were
taken with respect to the reference frequencies of aqueous YCl$_{3}$ and
LiCl solutions at $\nu =15634.15$ kHz and $\nu =123994.0$ kHz .

Let us first compare the effects of Li and Zn on the magnetism of the doped
CuO$_{2}$planes. In the underdoped regime, $^{89}$Y NMR established that Zn
induces local moments on its four near neighbor (n.n.) Cu sites\cite
{MahajanPRL94}. We have performed similar experiments on deoxidized Li
substituted samples. A typical spectrum is presented in the inset of Fig.1.
As in the Zn case (fig.1 of \cite{MahajanPRL94}), three lines are resolved.
The low frequency line is attributed to the Y n.n. sites, the second line to
further shells of neighbors, and the main line to more distant Y sites. The
temperature dependence of the shifts of these lines is reported in Fig.1 for
both Li, Zn and the pure compound. The Y NMR shift is proportional to the
magnetic susceptibility of the eight surrounding Cu sites of the two
adjacent CuO$_{2}$ planes. The striking similarity of the spectra and the T
dependences of the three lines for Zn and Li must result from a common local
magnetic perturbation, which is likely characteristic of all closed shell
spinless ion substitutions. In particular, the first n.n. shift exhibits a
1/T contribution in both cases, providing local evidence for the presence of
nearby moments. The strong similarity between Zn and Li would not be
expected if the extra hole given by Li$^{+}$ were localized on the n.n.
oxygen orbitals (as suggested in \cite{PoilblancPRL94} or \cite{Haase}).
Such a hole would carry a spin and make a markedly different contribution to
the Y n.n. shift. The hyperfine coupling of the Y nucleus to the oxygen
orbitals is considerably larger than to the Cu orbitals, thus the shift
would be much larger than observed. Our data suggest that any additional
hole delocalizes in the band of carriers of the CuO$_{2}$ planes. Also,
there are some indications that the introduction of Li is accompanied by an
oxygen loss to maintain overall electroneutrality, as shown by iodometric
titration \cite{Sauv}. In our dilute samples, these effects are not large
enough to be detected. The delocalization of the Li hole contrasts with
results in Li-doped antiferromagnet La$_{2}$CuO$_{4}$ which remains
insulator \cite{Sarrao}, underlining the fact that the metallic and the
antiferromagnetic regimes respond differently to the presence of an in-plane
hole. In Fig.1, the small difference between the amplitude of the n.n. Y
shift for Zn and Li corresponds to a Li induced moment with p$_{eff}$ 20\%
larger than for Zn. This could be due to a difference of doping level
between the Zn and Li samples\cite{NoteDopage}. In all, these results show
that Li$^{+}$ and Zn$^{2+}$ have the same effect on local magnetism in the
CuO$_{2}$ planes.

One can wonder whether their influence on scattering and superconductivity
is also similar. The depression of T$_{c}$ by Li is apparently much smaller
than for Zn, if we consider the nominal Li concentration $x_{n}$. As the
dominant effects on T$_{c}$ originate only from in-plane substitutions, such
a conclusion cannot be drawn because the actual distribution of Li in the
compound is not known. One of the great advantages of the present
experiments is to provide a quantitative estimate of the concentration $%
x_{plane}$ of Li in the planes, through measurements of the intensities of $%
^{89}$Y n.n. and $^{7}$Li NMR lines which are proportional to the number of
Y or Li contributing nuclei. The relative intensity of the first n.n. line
to the whole $^{89}$Y NMR spectrum scales with $4x_{plane}$ in the dilute
limit. While Zn was found to fully substitute in the planes ($%
x_{plane}\simeq 1.5x_{n}$)\cite{MahajanPRL94}, here we find that about half
of the nominal Li substitutes in-plane ($x_{plane}=0.85(\pm 0.2)$$x_{N}$).
The $^{7}$Li NMR intensity is another independent estimate of the relative
number of in-plane Li which we used to compare samples, confirming the
previous results from $x_{N}=1$ to $8\%$. Using the experimentally
determined plane Li concentration, we find that Li and Zn induce similar
reductions of T$_{c}$ versus $x_{plane}$ (see Fig.2). For optimal doping,
the effects of Li ($6\pm 1$\ K/\%\ ) and Zn ($7.1\pm 0.5$\ K/\%) are the
same within error bars. In underdoped samples, the slight difference between
Li and Zn might again originate from a slight doping difference \cite
{NoteDopage}. So, we conclude that Li and Zn produce {\it quantitatively}
the same effects both on T$_{c}$ and on the local magnetic structure
associated with the defect. The independence of these effects on the valence
of the impurity indicates that the scattering cross section responsible for
the decrease of T$_{c}$ is dominated, not by the relative charge of the
impurity but rather by the spinless perturbation created in the correlated
magnetic system.

Having established the existence of induced moments in underdoped Li
substituted compounds, we now proceed to study the evolution of these
moments with doping and concentration. The $^{89}$Y n.n. shift cannot be
used at optimal doping because the corresponding NMR line can no longer be
distinguished from the main line. In the case of Zn, only long range
magnetic effects of the impurity or macroscopic measurements could be used
to demonstrate the existence of local moments. A new possibility is offered
here by $^{7}$Li NMR\cite{Sauv}. Because $^{7}$Li has a spin I=3/2, the NMR
transitions between Zeeman levels $-3/2\longleftrightarrow -1/2$, $%
-1/2\longleftrightarrow 1/2$, and $1/2\longleftrightarrow 3/2$ will be split
by an electric field gradient (EFG) at the Li site. For our aligned samples,
with H//c, the quadrupolar spectrum of $^{7}$Li is composed of three narrow
lines separated by the quadrupolar frequency $\nu _{c}$ proportional to the
EFG along c. This gradient parameter $\nu _{c}$ ranges from 40 to 57 kHz
depending on the doping level\cite{bobroffLongPapier}, and similar
quadrupolar frequencies are inferred from the powder pattern in the
perpendicular orientation. Such small values of the quadrupolar frequencies
relative to the Larmor frequency simplify significantly the analysis of the
NMR spectrum. In particular, the position of the $-1/2\longleftrightarrow 1/2
$ transition is determined only by the magnetic environment of Li. The shift
of this transition is reported versus temperature for two dopings and two
concentrations of Li in Fig.3. In the underdoped compound, a clear 1/T
behavior is observed as is found with the shift of the n.n. $^{89}$Y line.
Thus $^{7}$Li NMR also senses the Curie behavior on its four n.n. Cu
analogous to the $^{27}$Al NMR in La$_{2}$SrCuO$_{4}$\cite{IshidaPRL96}. The
hyperfine coupling $A_{hf}$ responsible for the Li shift is nearly
isotropic, since the shift is almost identical for H//c and H$\perp $c. We
use $K_{c}=4A_{hf}\chi /N_{A}\mu _{B}$ with $\chi /N_{A}$ the atomic Curie
susceptibility of one induced n.n. Cu moment. We can estimate this
susceptibility by scaling the measured value p$_{eff}=1\mu _{B}$ for Zn by
the ratio 1.2 obtained above from the comparison of the n.n. $^{89}$Y
shifts. We obtain a value $A_{hf}^{iso}=2.5kOe/\mu _{B}$ which can be
explained by a super-exchange mechanism between Li nucleus and the four n.n.
Cu 3d$_{x^{2}-y^{2}}$. This mechanism is analogous to that proposed by Mila
and Rice for the hyperfine couplings between n.n. Cu in CuO$_{2}$ planes 
\cite{MilaRice}. In fact, using an LCAO computation as in \cite{MilaRice},
we find $A_{hf}=1.6$ to $3.9kOe/\mu _{B}$ \cite{bobroffLongPapier}, which is
compatible with the experimental value. It is apparent in Fig.3 that the
sensitivity of the Li probe is high enough to allow precise measurement of
the temperature dependence of the n.n. susceptibility $\chi $ even in
optimally doped samples. As already observed for Zn \cite
{Mendels99,MahajanPRL94}, this susceptibility falls off strongly with
increased doping, but still has a significant T-dependence at optimal
doping. This motivated us to investigate an overdoped YBaCuO system obtained
by Ca substitution for Y at maximal oxygen content \cite{TallonPRB95}. The
shift still varies substantially, increasing from $189$ to $287\pm 4$ ppm
between $450$ K and $100$ K, confirming that, {\em even in overdoped
materials, there is still a substantial induced moment}. However, as can be
seen in Fig.3, the T dependence no longer follows a 1/T law at optimal
doping. This is apparent in Fig.4 where the shift is seen to progressively
deviate from 1/T with increasing doping. Within experimental accuracy, all
the data can be fitted with a C/(T+$\Theta $) law. Similar T dependence was
observed in optimally doped Al substituted LaSrCuO \cite{IshidaPRL96}. The
present data permits us to study the $x_{plane}$ and $y$ dependence of C and 
$\Theta $. For the dilute Li concentrations $x_{plane}<2\%$ studied here,
the susceptibility and therefore $\Theta $ are found to be independent of
impurity concentration as seen in Fig.3, implying that $\Theta $ is not due
to interactions between induced moments of different Li sites. The variation
of $\Theta $ with hole doping is represented in the inset of Fig.4, where $%
\Theta $ increases markedly for $y>6.8$ and reaches a value as high as $226$
K for the Ca overdoped sample. It has been proposed that $\Theta $
originates from antiferromagnetic interaction between adjacent n.n. Cu sites
of the impurity \cite{IshidaReply}. However, increased doping is known to
reduce the strength of the magnetic correlations and should, in this
scenario, also reduce $\Theta $, in contrast to what we observe. $\Theta $
represents a new energy scale which evolves sharply with hole doping.

This 1/(T+$\Theta $) susceptibility is analogous with the Kondo effect
observed for dilute alloys with magnetic impurities. In these systems, the
coupling of the impurity spin with the band induces a screening of the local
moment below T$_{K}$, which yields $\chi \propto p_{eff}^{2}/($T$+$T$_{K})$.
Though, here it is a nonmagnetic impurity which induces such a behavior in a
correlated electronic band. Such a ``Kondo'' scenario has been proposed by
Nagaosa and Lee to explain the effects of Zn on the resistivity \cite
{NagaosaPRL97}. They argue, as have others\cite{Gabay}, that at high doping,
the moment is screened, which is qualitatively consistent with our observed
increase of $\Theta $ with doping. The value of C obtained from the fits is
found to vary between $5000$ and $8500$ kHz.K, $\pm 1500$kHz.K. Thus, the
moment value p$_{eff}\propto \sqrt{\text{C}}$ does not evolve strongly with
hole doping in contrast with $\Theta $, as also observed in the classical
Kondo effect. This apparently contrasts with the strong decrease of p$_{eff}$
with doping found in macroscopic susceptibility measurements on Zn
substituted YBaCuO \cite{Mendels99}. However, in such measurements, the
magnetic contribution of Zn is so small at optimal doping that it can be
extracted accurately only for large Zn concentrations ($x_{plane}=6\%$). In
this regime, $1/$T behavior was apparent, and its reduced amplitude was
attributed to a reduction in the net moment. The data on Zn and Li could be
reconciled if $\Theta $ is progressively suppressed with increasing impurity
concentration, as was observed in dilute alloys such as Cu:Fe \cite
{TholencePRL70}, where at high impurity concentrations, interaction between
the moments restores a 1/T behavior.

In conclusion, the influence of Li and Zn on the magnetic and T$_{c}$
properties is governed by the presence of magnetic correlations. In
particular, the strong impurity scattering appears to be connected with the
absence of spin and is insensitive to charge. Such a spinless site induces
moments in the correlated electronic system, which persist into the
overdoped regime, clearly indicating that in-plane correlations are not
suppressed and that pure uncorrelated Fermi Liquid behavior has not yet
developed. While these moments display a Curie law in the underdoped regime,
a Kondo-like temperature $\Theta $ appears and increases sharply with
doping. It is very interesting to consider possible connection of this
magnetic screening energy to the pseudogap and T$_{c}$ energy scales which
both vary considerably in the same range of doping. The analogy with a Kondo
effect is at the present stage phenomenological. Detailed experimental
investigations of other normal state properties such as transport are needed
to understand whether this analogy proves relevant.

We wish to acknowledge F.\ Mila for fruitful discussions.

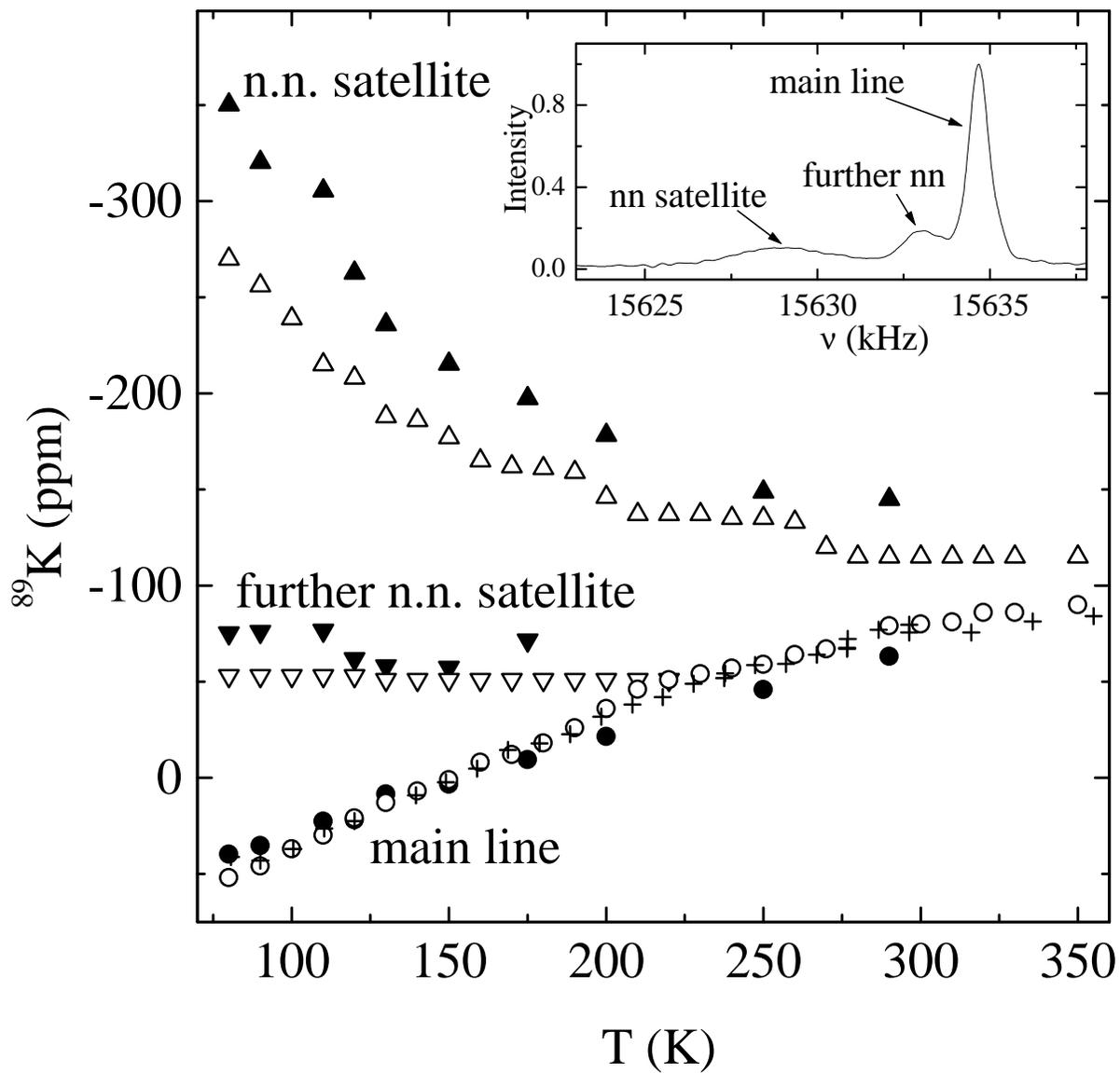
\begin{figure}[tbp]
\caption[1]{$^{89}$Y\ NMR shift for H$\bot $c, for pure (cross), Zn\ $%
x_{n}=0.5\% $ (open symbols, taken from \protect\cite{MahajanPRL94}) and Li\ 
$x_{n}=1\%$ (closed symbols) underdoped YBa$_{2}$Cu$_{3}$O$_{6.6}$ samples.
The three datasets correspond to the shifts of the three lines of the
spectrum. A typical one taken at 100K for the Li sample is displayed in the
inset.}
\label{fig.1}
\end{figure}

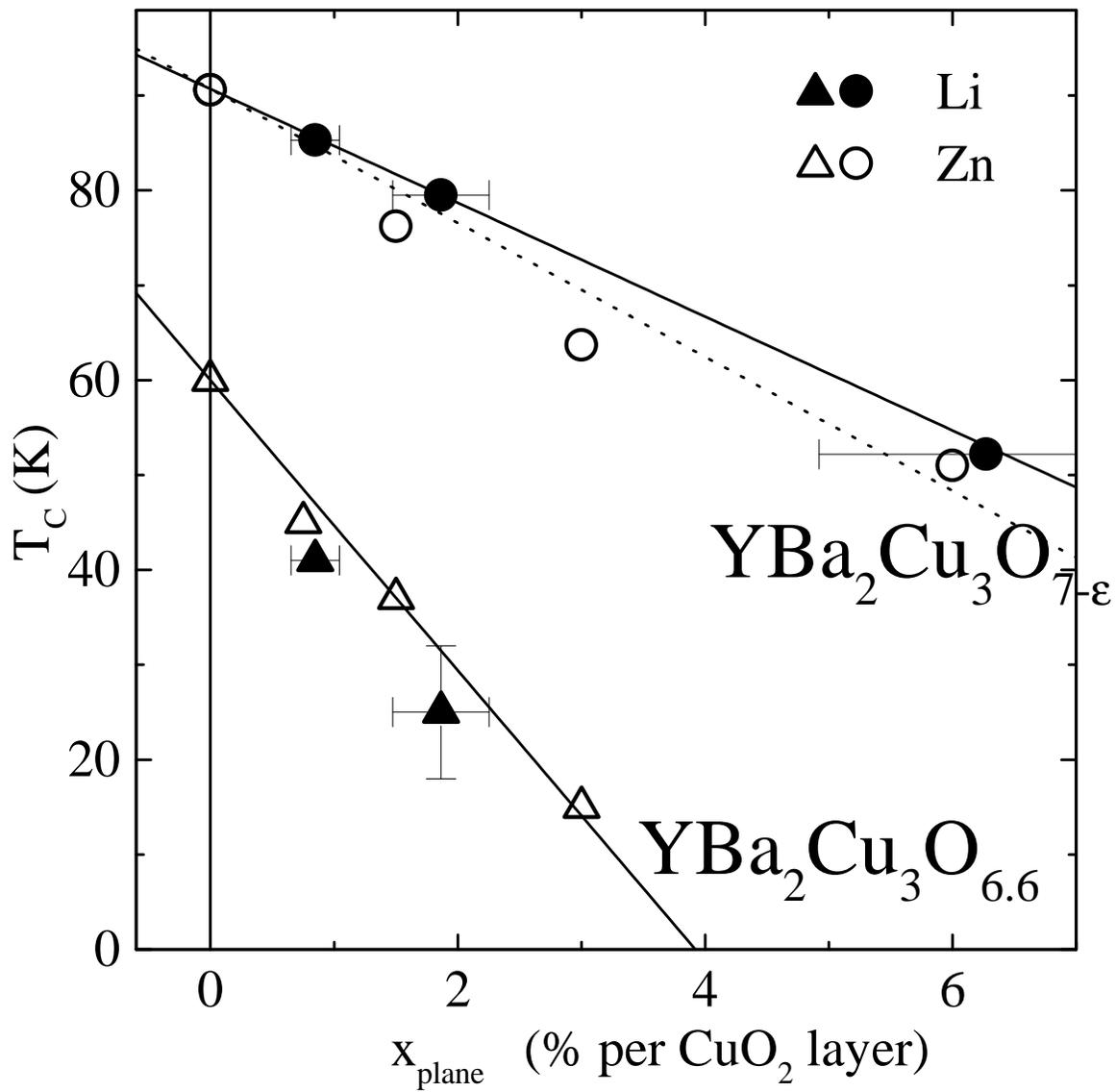
\begin{figure}[tbp]
\caption[2]{ Variation of T$_{c}$ versus the concentration $x_{plane}$ of Zn
or Li per CuO$_{2}$ layer, both for underdoped and optimally doped
compounds. $x_{plane}$ is deduced from the intensities of the n.n. $^{89}$Y
and $^{7}$Li NMR lines.}
\label{fig.2}
\end{figure}

\begin{figure}[tbp]
\caption[3]{$^{7}$Li NMR frequency for H//c in YBa$_{2}($Cu$_{1-x}$Li$%
_{x})_{3}$O$_{6+y}$ for $x_{plane}=0.85\%$ (closed symbols) and $%
x_{plane}=1.86\%$ (open symbols) for underdoped $y=0.6$ (circles) and
optimally doped (diamonds) compounds. Deoxydation is performed on samples
from the same batch.}
\label{fig.3}
\end{figure}

\begin{figure}[tbp]
\caption[4]{$^{7}$Li NMR frequency for H//c in YBa$_{2}($Cu$_{99\%}$Li$%
_{1\%})_{3}$O$_{6+y}$ for $x_{plane}=0.85\%$ and different oxygen contents $%
y $ on samples taken from the same batch. The Y$_{80\%}$Ca$_{20\%}$Ba$_{2}($%
Cu$_{99\%}$Li$_{1\%})_{3}$O$_{6+y}$ compound corresponds to an overdoped T$%
_{c}=63$K sample. Black lines correspond to fit in (T+$\Theta $)$^{-1}$
which are also shown in fig.3. $\Theta $ is plotted versus oxygen content $y$
in the inset. \ The doping level is determined through the main line $^{89}$%
Y NMR shift calibrated with pure compounds with different doping levels.}
\label{fig.4}
\end{figure}


\begin{references}
\bibitem{NoteDopage}  Near $y=6.6$, the Y\ shift does not allow to
distinguish between doping levels different by less than $\Delta y=0.05$.
\bibitem{bobroffLongPapier}  Full detailed analysis will be published
elsewhere.
\end{references}
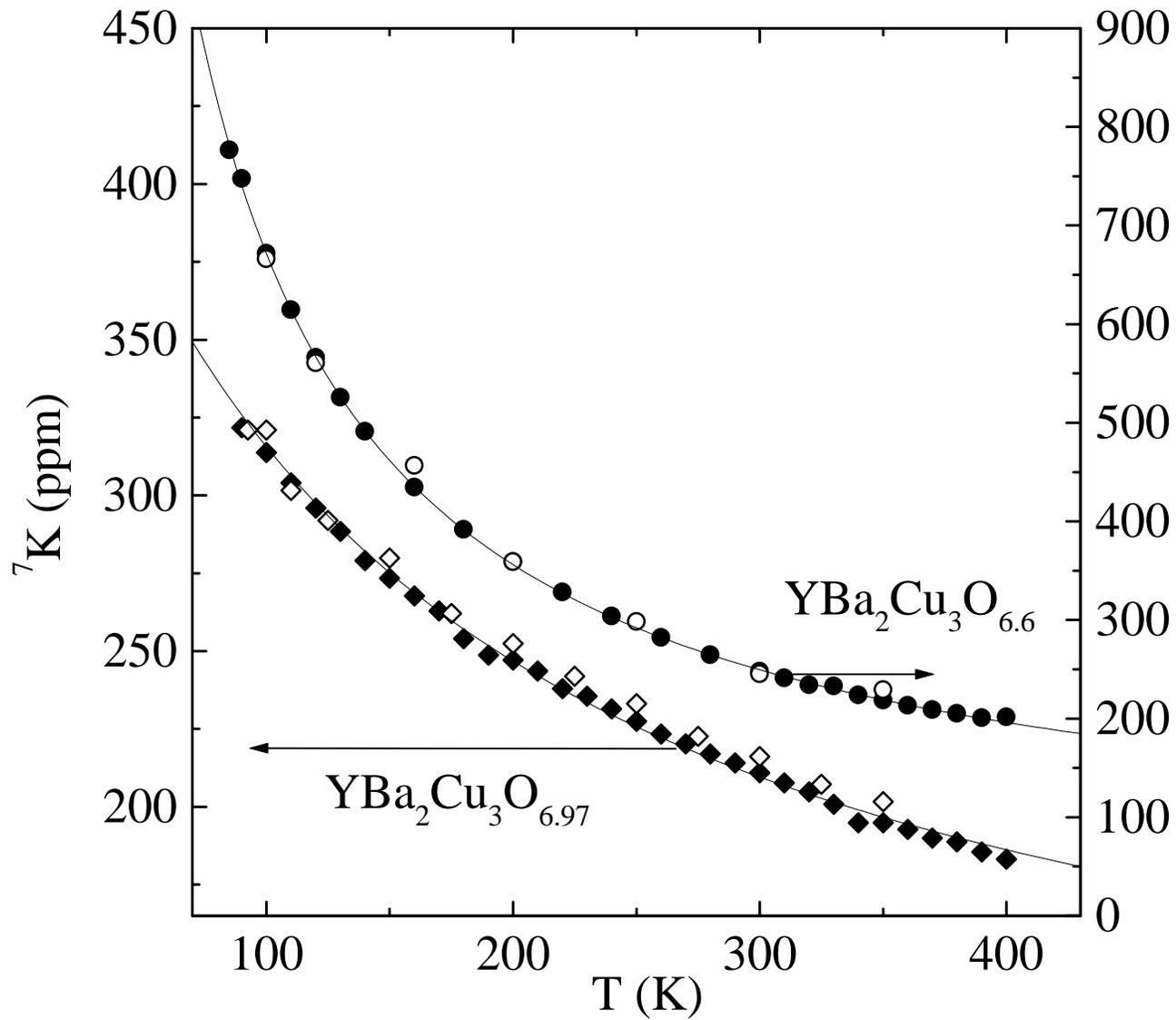
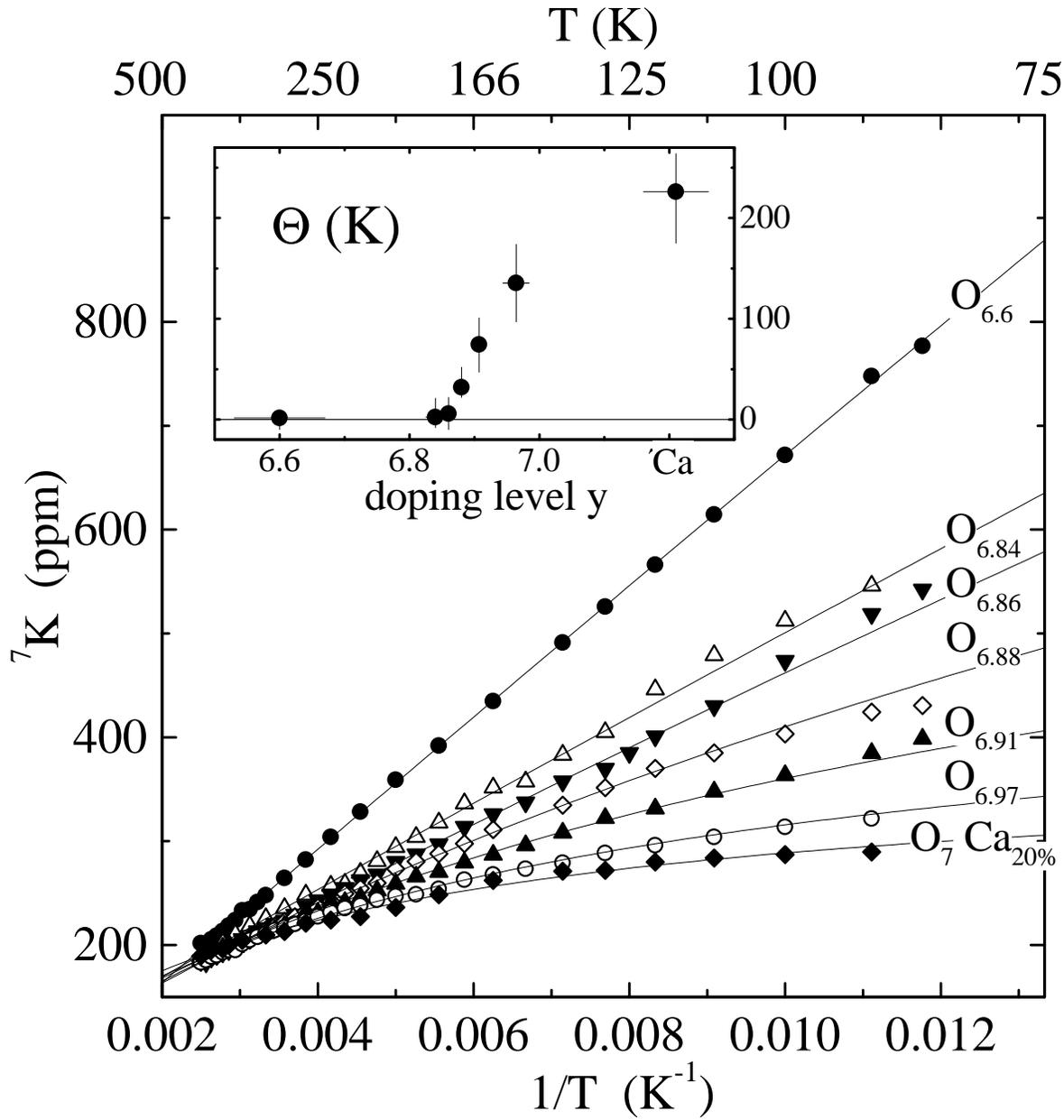
\end{document}